%% file: jylha_dngJAP.tex
\documentclass[preprint,aps,altaffilletter]{revtex4}

\usepackage{graphicx}% Include figure files
\usepackage{dcolumn}% Align table columns on decimal point
\usepackage{bm}% bold math
\usepackage{pst-all}
\usepackage{subfigure}
\usepackage{multirow}

\begin{document}

%\preprint{APS/123-QED}

%\renewcommand{\altaffilletter@sw}{\alph{1}}

\title{Modeling of Isotropic Backward-Wave Materials Composed of Resonant
Spheres}

\author{L. Jylh\"{a}}
 \email{Liisi.Jylha@tkk.fi}
\altaffiliation{Electromagnetics Laboratory, P.O. Box 3000, Helsinki University of Technology, Finland }
\author{I. Kolmakov}%
%\email{i_k@rambler.ru}
\altaffiliation{Radio Laboratory / SMARAD, Helsinki University of Technology, P.O. 3000,
 FI-02015 TKK, Finland and Microwave Microelectronics Laboratory, Electrotechnical
University, Prof. Popov 5, 197376 St. Petersburg, Russia}
\author{S. Maslovski}%
\email{Stanislav.Maslovski@tkk.fi}
\altaffiliation{Radio Laboratory / SMARAD, Helsinki University of Technology, P.O. 3000,
 FI-02015 TKK, Finland}
\author{S. Tretyakov}
 \email{Sergei.Tretyakov@tkk.fi}
\altaffiliation{Radio Laboratory / SMARAD, Helsinki University of Technology, P.O. 3000,
 FI-02015 TKK, Finland}

\date{\today}% It is always \today, today,
\begin{abstract}
A possibility to realize isotropic artificial backward-wave
materials is theoretically analyzed. An improved mixing rule for
the effective permittivity of a composite material consisting of
two sets of resonant dielectric spheres in a homogeneous
background is presented. The equations are validated using the Mie
theory and numerical simulations. The effect of a statistical
distribution of sphere sizes on the increasing of losses in the
operating frequency band is discussed and some examples are shown.
\end{abstract}

%Use showkeys class option if keyword
\maketitle

\section{Introduction}

Most of the known realizations of artificial backward-wave
materials with negative parameters (also called  Veselago media,
negative index materials, and double-negative materials)
\cite{shelby,parazzoli} are highly anisotropic composites. The
usual design utilizes arrays of thin conductors and arrays of
split rings. In addition to anisotropy and, in most cases, even
bi-anisotropy of these structures, they exhibit strong spatial
dispersion due to the presence of long conductors that support TEM
modes \cite{belov}. However, for many applications, isotropic
materials are required or preferable. To provide isotropic
response with negligible spatial dispersion, one can possibly use
regular or random arrays of resonant particles that have both
electric and magnetic resonances. One possibility is to use
$\Omega$-shaped metal inclusions. This opportunity has been
explored theoretically in \cite{simovski2003,simovski2004} and
experimentally in \cite{verney}. The use of racemic mixtures of
chiral particles is also a possibility \cite{modeboo}. However, in
both these solutions the inclusions are bi-anisotropic particles,
and the material behaves as an effectively isotropic medium only
at the macroscopic scale, if there are enough particles
distributed in a proper fashion. This means that the use of
isotropic particles is clearly preferable.

The idea of an isotropic backward-wave material constructed from
small isotropic spheres in a dielectric background was presented
in \cite{holloway}. In that paper, the use of two sets of spheres
was proposed. In one set,  the spheres are made of a
high-permittivity dielectric, and in the other set they are made
of a high-permeability magnetic material. Here, resonant
dielectric spheres provide effective negative permittivity, and
resonant magnetic spheres provide negative permeability. In paper
\cite{vendik2004}, a more practical design was suggested, such
that all spheres are made of a dielectric material, but there are
two sets of spheres with different radii. The dielectric constant
of spheres is larger than that of the background. Then the
wavelength inside the sphere is comparable to the diameter of the
sphere and simultaneously the wavelength outside the sphere is
large compared to the sphere. By combining two sets of spheres
with suitable radii, one set of spheres is in the magnetic
resonance mode and the other set is in the electric resonance
mode.

\begin{figure}[h!]
\begin{center}
\input{fig1.tex}
\caption{A composite construct with two sets of dielectric spheres in a dielectric background. If the wavelength is much larger than the radii of the spheres and the distance between them, the material can be modeled with effective
permittivity $\epsilon_{\rm eff}$ and permeability $\mu_{\rm eff}$.}
\label{illustration}
\end{center}
\end{figure}
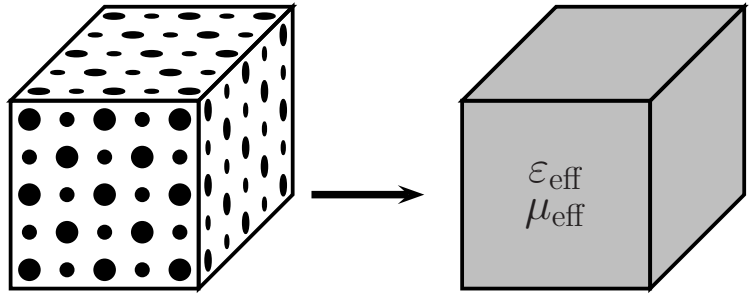

The effective permittivity and permeability can be calculated
using effective medium models if the wavelength outside spheres is
large compared to sphere diameters and the polarizability of
spheres is known. An illustration  of the homogenization  problem
is shown in Fig.~\ref{illustration}. Earlier studies
\cite{holloway,vendik2004} are based on effective medium theories
for a composite material consisting of spheres which resonate
either in the electric or magnetic mode, as presented in
\cite{lewin}. In Lewin's model, spheres are assumed to resonate
either in the first or second resonance mode of the Mie theory. In
\cite{holloway,vendik2004} Lewin's equations were applied directly
to the system of two sets of spheres, and the electric
polarizability of spheres in the magnetic resonant mode was not
taken into account. However, electrical properties of these spheres can
have a significant effect on the effective permittivity of the
composite. This affects especially the low frequency limit, which
then approaches to the classical Maxwell-Garnett mixing rule. In
this paper we present an effective medium model, which takes this
effect into account. Equations are validated both analytically and
numerically. Scattering from a single sphere is calculated both
analytically from the full Mie theory and numerically. It is shown
numerically, that there exists a backward wave in the frequency
band, where the effective medium theory predicts it.

The electric and magnetic polarizabilities of spheres can be
calculated using the Mie theory \cite{bohren}. The scattered
electromagnetic field is expressed as an infinite series of vector
spherical harmonics $M_n$ and $N_n$, each weighted by appropriate
amplitude coefficients $a_n$ and $b_n$. These represent the normal
electromagnetic modes of the spherical particle. In general, the
field is a superposition of normal modes of fields. This can be
interpreted as a superposition of fields generated by electric and
magnetic dipoles, and multipoles. When the permittivity of the spheres is much
larger than the permittivity of the environment,
around the eigenfrequencies of
the first two modes $a_1$ and $b_1$, the dipole fields dominate.
Therefore frequency dependent polarizabilities can be introduced:
The problem is still quasistatic, because the field outside the
sphere can be modeled with a dipole field. For each order $n$
there are two distinct types of modes: transverse magnetic (TM)
and transverse electric  (TE) modes. When $n=1$, these modes
represent the field scattered by a magnetic dipole (TM) and by an
electric dipole (TE). In the low frequency limit, these frequency
dependent polarizabilities approach static polarizabilites.

The statistical size distribution of spheres, which is the actual
case when this material is realized, is also analyzed. The size
distribution causes increasing losses in the frequency band, where
backward waves can exist.

\section{Mixing theory}

In  this section, the Clausius-Mossotti (Maxwell-Garnett) mixing
relation and polarizabilities of spheres near the first two Mie
resonance modes are presented. The material of spheres and the
background are assumed to be dielectric and non-magnetic. Footnote
$e$ corresponds to spheres in the electric resonance mode and $m$
to the magnetic resonance mode.

The effective permittivity $\epsilon_{\rm eff}$ for a material
with two types of inclusions having two different electric
polarizabilities can be calculated from the generalized Claussius
Mossotti relation \cite{ari}:
\begin{equation}
\frac{\epsilon_{\rm eff}-\epsilon_b}{\epsilon_{\rm eff}+2\epsilon_b}=\frac{n_e \alpha_e}{3\epsilon_b}+\frac{n_m \alpha_m}{3\epsilon_b}
\label{eps_eff}
\end{equation}
where $n_m$ and $n_e$ are the number of spheres per unit volume in
the magnetic resonance and in the electric resonance,
respectively, $\alpha_m$ and $\alpha_e$ are the electric
polarizabilities of spheres in the magnetic resonance and in the
electric resonance mode. In \cite{holloway,vendik2004} it was
assumed that $\alpha_m=0$. Neglecting $\alpha_m$ leads to an error
which is usually small near the resonant frequency, but the error
out from the resonant frequency can be quite significant. However,
the error near the resonant frequency increases as the electrical
contrast between inclusions and the environment increases. Also
the low frequency limit for the effective permittivity is
incorrect, because the remaining static electric polarizability of
spheres in the magnetic resonance modes is not taken into account.
The effective permeability can be obtained from (\ref{eps_eff}) by
replacing $\epsilon$ with $\mu$ and the electric polarizability
$\alpha$ with the magnetic polarizability $\beta$. We can assume
that the magnetic polarizability of spheres in the electric
resonance $\beta_e=0$, because materials are non-magnetic:
$\mu_b=\mu_i=1$. Then the magnetic polarizability arises only from
resonant phenomena, and there is no remaining static value.
Therefore, the effective permeability $\mu_{\rm eff}$ reads:
\begin{equation}
\frac{\mu_{\rm eff}-1}{\mu_{\rm eff}+2}=\frac{n_m \beta_m}{3\mu_b}
\label{mu_eff}
\end{equation}
where $\beta_m$ is the magnetic polarizability of spheres in the magnetic resonance mode.

Frequency dependent polarizabilities $\alpha$ and $\beta$ for a
sphere with radius $r$ are calculated using the Mie theory.
Polarizabilitiy $\alpha_e$ corresponds to the electric polarizability
with $r=r_e$, $\alpha_m$ to that with $r=r_m$, and $\beta_m$
corresponds to the magnetic polarizability with $r=r_m$.
 The coefficients for the spherical harmonics in
the first two Mie resonance modes (where $n=1$) are
\cite{lewin,bohren}:
\begin{equation}
a_1=j\frac{2}{3}(k_0^2\mu_b \epsilon_b)^{3/2} \frac{\mu_b -\mu_i F(\Theta)}{2\mu_b+\mu_i F(\Theta)}r^3
\label{a1}
\end{equation}
\begin{equation}
b_1=j\frac{2}{3}(k_0^2\mu_b \epsilon_b)^{3/2} \frac{\epsilon_b -\epsilon_i F(\Theta)}{2\epsilon_b+\epsilon_i F(\Theta)}r^3
\label{b1}
\end{equation}
where
\begin{equation}
F(\Theta)=\frac{2(\sin\Theta-\Theta \cos\Theta)}{(\Theta^2-1)\sin\Theta+\Theta\cos\Theta}, \hspace{1cm}
\label{F}
\end{equation}
\begin{equation}
\Theta=k_0r\sqrt{\epsilon_i \mu_i}
\label{theta}
\end{equation}
The scattering parameter $S_1$ \cite{bohren} for a sphere, when
only the order $n=1$ modes are excited reads:
\begin{equation}
S_1=\frac{3}{2}(a_1+b_1)
\label{S1}
\end{equation}
This corresponds to scattering from electric and magnetic dipoles: $S_1=S_{1e}+S_{1m}$.
Scattering parameter $S_{1e}$ for an electric dipole can be written in terms of the
electric polarizability $\alpha$:
\begin{equation}
S_{1e}=\frac{jk^3\alpha}{4\pi \epsilon_b}
\label{S1e}
\end{equation}
The scattering parameter for a magnetic dipole is
\begin{equation}
S_{1m}=\frac{jk^3\beta}{4\pi \mu_b}
\label{S1m}
\end{equation}
By combining equations ($\ref{a1}$--$\ref{S1m}$) with $\mu_b=1$
and $V_{\rm sphere}=4\pi r^3/3=f/n$, where $f$ is the volume
fraction of spheres $f=V_{\rm sphere}/V_{\rm tot}$, the
polarizabilities of electric and magnetic dipoles can be written
as:
\begin{equation}
\alpha=4\pi r^3\epsilon_b\frac{\epsilon_b-\epsilon_i F(\Theta)}{2 \epsilon_b+\epsilon_i F(\Theta)}=\frac{3f\epsilon_b}{n}\frac{\epsilon_b-\epsilon_i F(\Theta)}{2 \epsilon_b+\epsilon_i F(\Theta)}
\label{alpha}
\end{equation}
\begin{equation}
\beta=4\pi r^3\frac{1-\mu_i F(\Theta)}{2 +\mu_i F(\Theta)}=\frac{3f}{n}\frac{1-\mu_i F(\Theta)}{2 +\mu_i F(\Theta)}
\label{beta}
\end{equation}
When these are substituted in the Clausius-Mossotti relations
(\ref{eps_eff}--\ref{mu_eff}) with $r=r_e$ for spheres in the
electric resonance and $r=r_m$ for spheres in the magnetic
resonance, effective medium models for a composite consisting of
two set of resonating spheres read:
\begin{equation}
\frac{\epsilon_{\rm eff}-\epsilon_b}{\epsilon_{\rm eff}+2\epsilon_b}={f_e}\left( \frac{2 \epsilon_b+\epsilon_i F(\Theta_e)}{\epsilon_b-\epsilon_i F(\Theta_e)} \right)+{f_m}\left( \frac{2 \epsilon_b+\epsilon_i F(\Theta_m)}{\epsilon_b-\epsilon_i F(\Theta_m)} \right)
\label{eps}
\end{equation}
\begin{equation}
\frac{\mu_{\rm eff}-1}{\mu_{\rm eff}+2}=f_m\left( \frac{2+F(\Theta_m)}{1-F(\Theta_m)} \right)
\label{mu}
\end{equation}
where $\Theta_e=k_0r_e\sqrt{\epsilon_i \mu_i}$ and
$\Theta_m=k_0r_m\sqrt{\epsilon_i \mu_i}$. These equations
(\ref{eps}, \ref{mu}) are similar to those given in
\cite{holloway,vendik2004} if the second term in the right hand
side in (\ref{eps}) is set to zero.

\begin{figure}[h!]
\begin{center}
\includegraphics[width=0.6\linewidth]{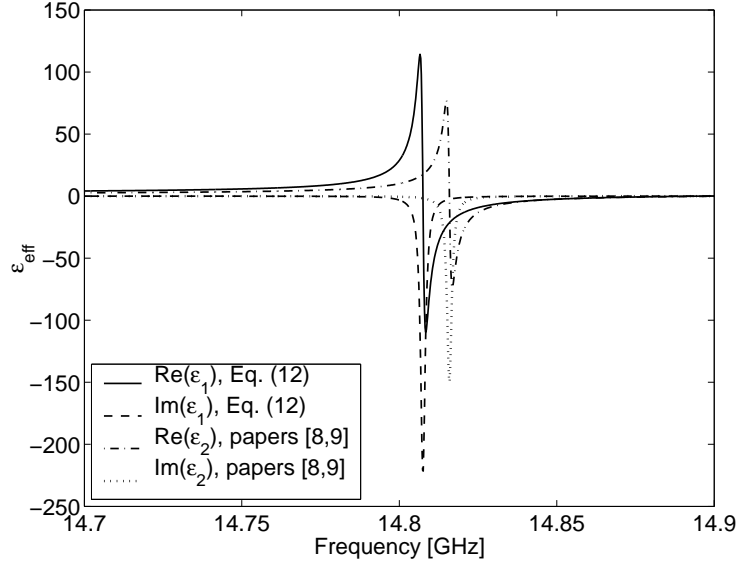}
\caption{The effective permittivity $\epsilon_1$ as a function of the frequency
calculated from Eq.~(\ref{eps})  compared to the effective permittivity $\epsilon_2$
 calculated without taking into account the electrical
polarizability of spheres in the magnetic resonance. In this case the
second term on the right-hand side of Eq.~(\ref{eps}) is zero.
$\epsilon_i=100(1-j1.25\cdot 10^{-4})$, $\epsilon_b=1$, $f_e=0.15$,
$f_m=0.15$, $r_e=3.18\cdot 10^{-3}$ mm, $r_m=2.18\cdot 10^{-3}$
mm.} \label{eps_corrected}
\end{center}
\end{figure}

In Fig.~\ref{eps_corrected} an example of effective permittivity
as a function of the volume fraction of spheres is shown. The
solid line represents $\epsilon_{\rm eff}$ given by
Eq.~(\ref{eps}), and the dashed line $\epsilon_{\rm eff}$ is
calculated using the method of \cite{holloway,vendik2004}, where
the electric polarizability of spheres in the magnetic resonance
mode is not taken into account. It can be seen that the resonant
frequency slightly shifts when the improved mixing equation is
used. In this case the old effective medium model overestimates
the effective permittivity out of the resonance mode.

\section{Statistical size distribution of spheres}

In practice, when a set of spheres is manufactured, the sphere
dimensions are statistically distributed because of production
inaccuracies.  The effect of a continuous statistical size
distribution on the real part of the effective permittivity and
the losses will be estimated next. The Clausius-Mossotti equation
for $K$ spheres in electric resonance and $N$ spheres in magnetic
resonance reads:
\begin{equation}
\frac{\epsilon_{\rm eff}-\epsilon_b}{\epsilon_{\rm eff}+2\epsilon_b}=\frac{1}{3\epsilon_b}\left( \sum_{k=1}^K n_{ek}\alpha_{ek}  + \sum_{n=1}^N n_{mn}\alpha_{mn} \right)
\label{eps2}
\end{equation}
By substituting the electric polarizabilites (\ref{alpha})
into equation (\ref{eps2}), with $r=r_e$ for the spheres in the electric resonance and $r=r_m$  for
the spheres in the magnetic resonance, we get:
\begin{equation}
\frac{\epsilon_{\rm eff}-\epsilon_b}{\epsilon_{\rm eff}+2\epsilon_b}=\sum_{k=1}^{K} \frac{f_{ek}}{G(\Theta_{ek})}+\sum_{n=1}^{N} \frac{f_{mn}}{G(\Theta_{mn})}
\label{eps4}
\end{equation}
This is a solution for the effective permittivity with two sets of
spheres. One set of spheres is in the magnetic resonance and the
other set is in the electric resonance. The effective permeability
can be calculated in a similar way.

Eq.~(\ref{eps4}) is a discrete summation. It can be interpreted
also as samples taken with the Dirac delta function from
continuous probability density functions $g_1(r_e)$ and
$g_2(r_m)$, which describe the particle size distributions of
spheres in the electric and magnetic resonances. They are
normalized as $\int_0^\infty g_1(r_e)\,dr_e= \int_0^\infty g_2(r_m)\,dr_m$=1.

Then the
effective permittivity for the continuous size distributions reads:
\begin{equation}
\frac{\epsilon_{\rm eff}-\epsilon_b}{\epsilon_{\rm eff}+2\epsilon_b}={f_e}\int_{0}^{\infty} \frac{g_1(r_e)}{G(\Theta_{e})}\, dr_e+{f_m}\int_{0}^{\infty} \frac{g_2(r_m)}{G(\Theta_{m})}\, dr_m
\label{eps_con}
\end{equation}
Here
\begin{equation}
G(\Theta_i)=\frac{\epsilon_b -\epsilon_i F(\Theta_i)}{2\epsilon_b + \epsilon_i F(\Theta_i)}
\end{equation}
functions $1/G(\Theta_e)$ and $1/G(\Theta_m)$ are integrable,
$f_e$ is the volume fraction of spheres in the electric resonance,
and $f_m$ is the volume fraction of spheres in the magnetic
resonance mode.

\begin{figure}[h!]
\begin{center}
\subfigure[$\sigma_e=\sigma_m=1\mu$m]{\label{mueps2}\includegraphics[width=0.42\linewidth]{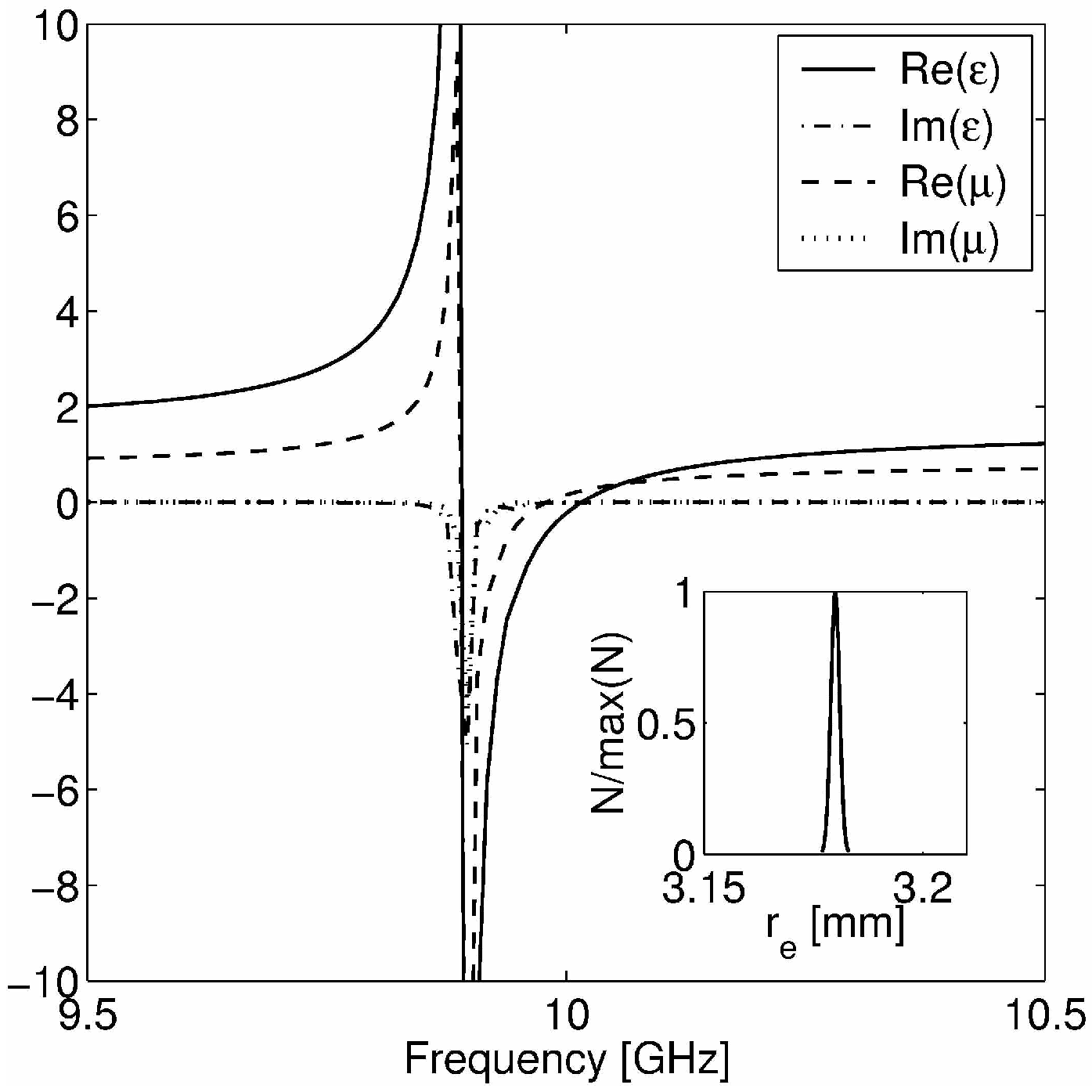}}
\subfigure[$\sigma_e=\sigma_m=10\mu$m]{\label{mueps3}\includegraphics[width=0.42\linewidth]{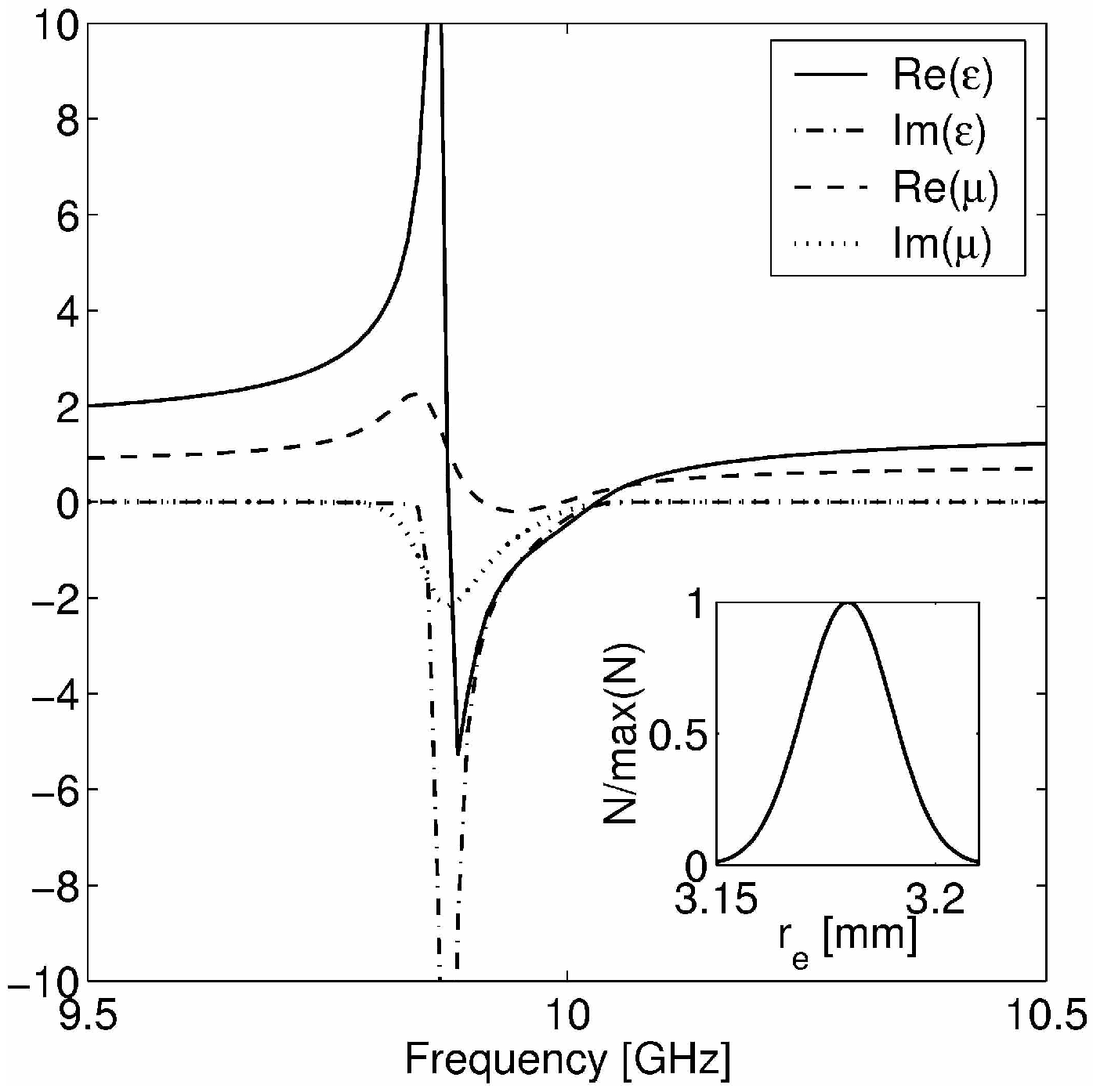}}
\caption{The effective permittivity as a function of the frequency
calculated using Eq.~(\ref{eps_con}) with
$\epsilon_i=44(1-j1.25\cdot 10^{-4})$, $\epsilon_b=1$ and filling
ratios $f_e=2\%$, $f_m=14\%$. On the left-hand side spheres are
normally distributed with the half-value widths
$\sigma_m=\sigma_e=10^{-5}$ and expectation values $r_e=3.18$ mm,
$r_m=2.28$ mm. The size distribution $N$ for
 $\sigma_e$ is also shown. On the right-hand side everything is the same,
expect the half value widths are $\sigma_e=\sigma_m=10^{-6}$.}
\end{center}
\end{figure}

An example on how a normal size distribution $N=1/(\sigma \sqrt(2
\pi)) \exp((r-\hat{r})^2/2)$ of spheres with half value widths
$\sigma=\sigma_e$ and $\sigma=\sigma_m$ and expectation values
$\hat{r}=r_e$ and $\hat{r}=r_m$ affects the effective material
parameters is presented in Figs.~\ref{mueps2} and \ref{mueps3}.
The half-value widths $\sigma_e=\sigma_m= 1$ $\mu$m
(Fig.~\ref{mueps2}) does not increase losses significantly, but
with the half-value widths of 10 $\mu$m (Fig.~\ref{mueps3}) the
imaginary part becomes large and $\mu_{eff}$ does not take
negative values.

We have not analyzed how the nonidealities in the lattice
structure will affect the scattering losses or how scattering
losses will increase because of the size distribution of spheres.

\section{Verification of the effective medium model}

In the original Lewin's paper \cite{lewin}, the validity limit of
effective medium equations was given as
$\lambda/|\sqrt{\epsilon_{\rm eff}\mu_{\rm eff}}|>10r$. This limit
is very strict, because it was derived by demanding reasonable
values when the filling fraction $f \rightarrow \infty$ in the
Clausius-Mossotti mixing equation. This requirement is not
necessary, because the mixing equation is anyway not valid when
$f\rightarrow \infty$.

In this paper, the frequency dependent
polarizabilities for spheres are given. To verify when these
polarizabilities are valid, one should study the complete Mie
theory expansion, where all the higher order modes are taken into
account. If the spheres can be replaced with frequency dependent
electric and magnetic dipoles (\ref{alpha}, \ref{beta}), then also
the Clausius-Mossotti effective medium models are valid when the
volume fraction of spheres is small. To study how well the
effective medium model works for high volume fractions of spheres,
one should first study if close separation of spheres causes
excitation of higher order resonance modes.

There are two
limitations in the model. The first limitation is that the
wavelength outside the spheres should be large compared to the
size of spheres, so that the material would work as an effective
medium and the spheres could be modeled by electric and magnetic
dipoles with frequency dependent polarizabilities. The second
limitation arises from the mixing theory: the filling fraction of
spheres should be small.

\begin{figure}[h!]
\centering
\subfigure[analytics, $Q_{\rm sca}$]{\label{Qsca}\includegraphics[width=0.4\linewidth]{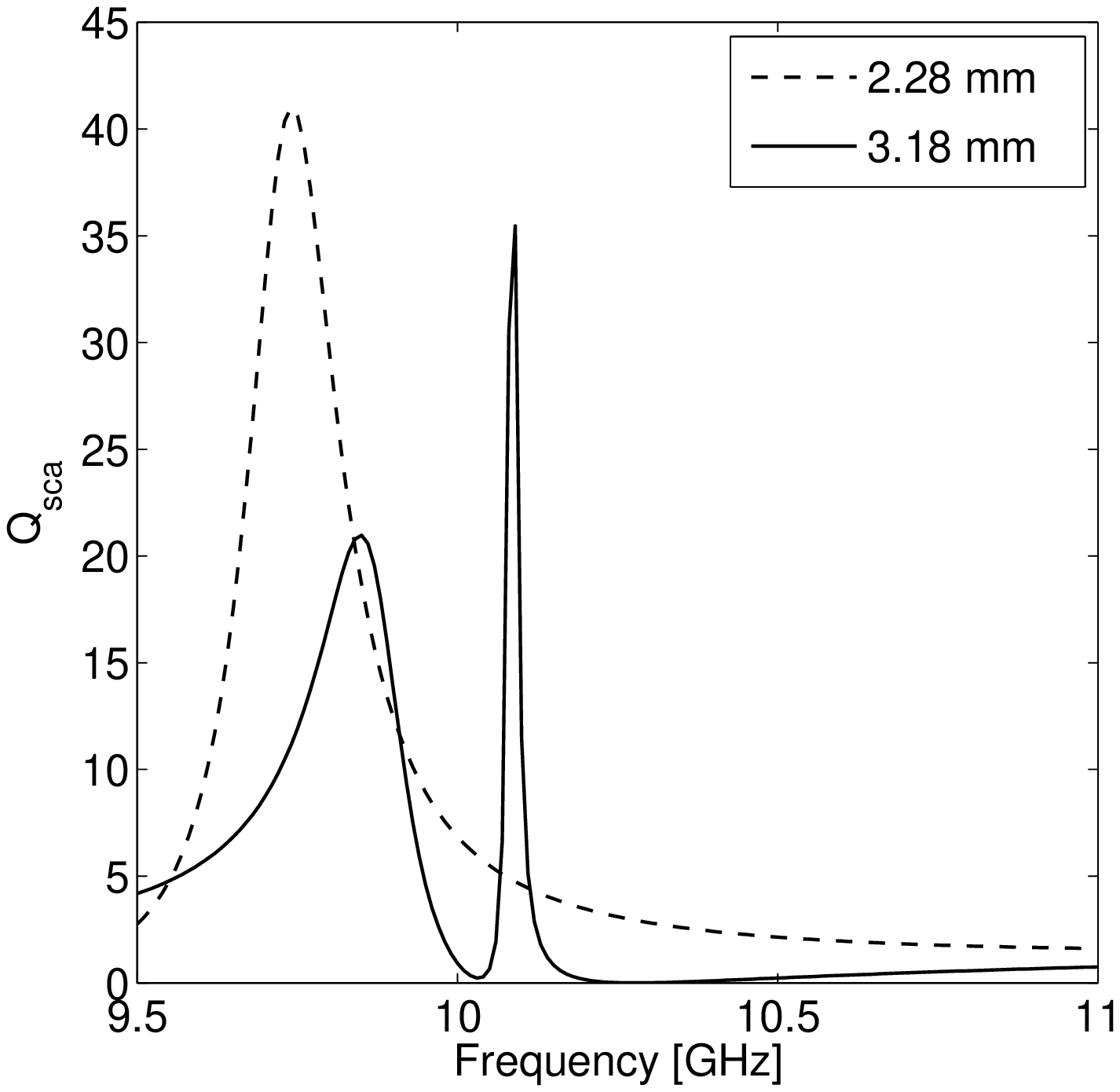}}
\subfigure[numerics, $|S_{11}|$]{\label{S11}\includegraphics[width=0.4\linewidth]{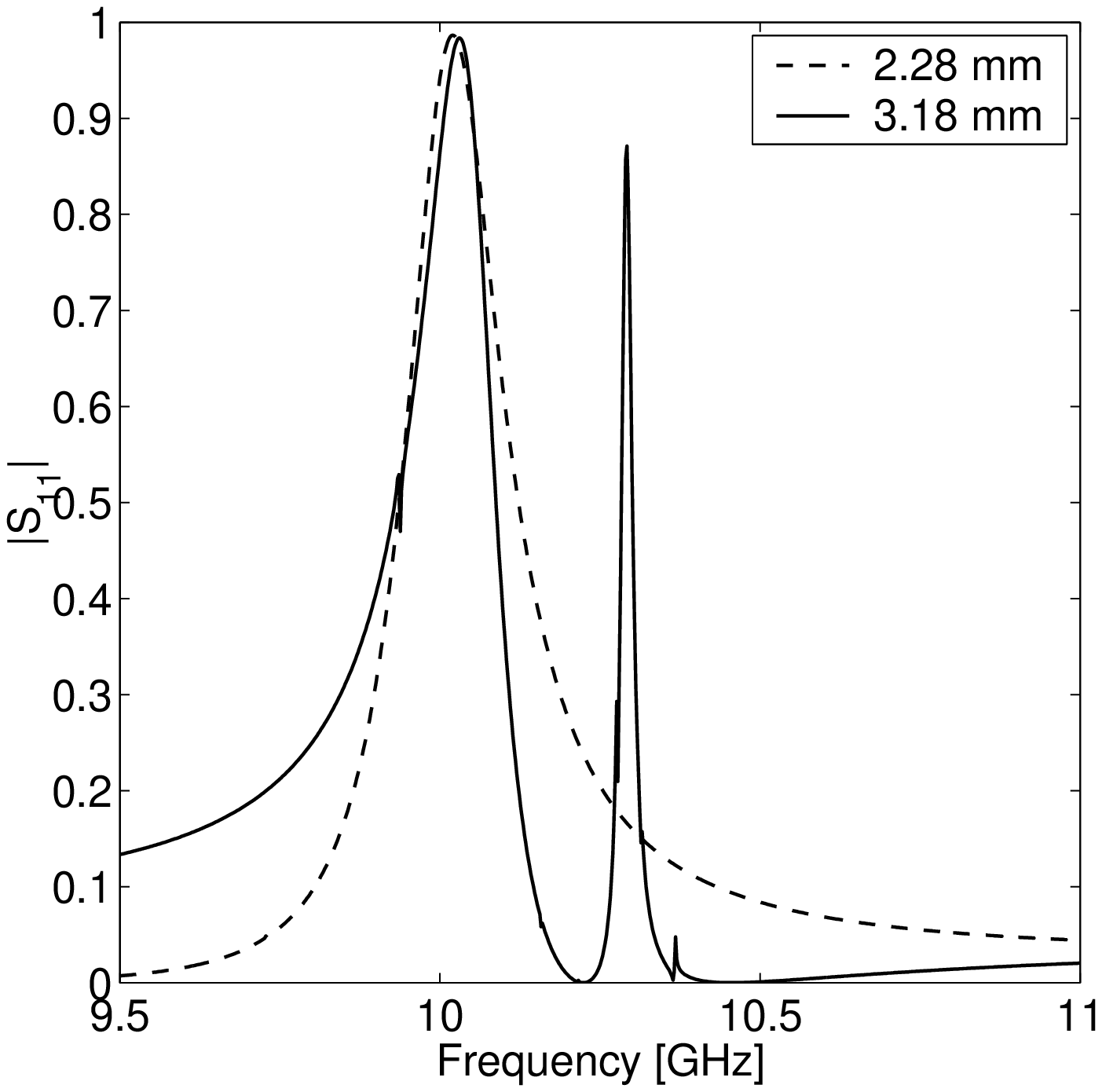}}
\caption{Analytically calculated RCS (a) and numerically calculated $|S_{11}|$ (b) for a sphere with $r_e=3.18$ mm and $\epsilon_i=44(1-j10^{-4})$.}
\end{figure}

The effective medium model is based on the assumption, that the
spheres resonate in the first (magnetic) and the second (electric)
resonance modes. This assumption can be tested using the Mie
theory. In Fig.~\ref{Qsca} the scattering cross section  ($Q_{\rm
sca}$) of a sphere with radii $r=r_e=3.18$ mm and $r=r_m=2.28$ mm
and $\epsilon_i=44(1-j10^{-4})$ is shown. Radii of spheres were
chosen so that the electric resonance and magnetic resonance occur
at approximately the same frequency. The smaller sphere resonates
clearly in the first resonance mode (magnetic), because no
additional peaks can be seen. The larger sphere has the second
resonant mode (electric) near the same frequency, but the third
resonant mode has its resonant frequency near the second mode.
This mode is not taken into account in the effective medium models
(\ref{eps}, \ref{mu}), which are based on the first two resonant
modes. However,  $Q_{\rm sca}$ goes nearly to zero between these
modes, which indicates that when the sphere is in the second
resonant mode, the effect of the third resonant mode can be
neglected. The third resonance mode sets an upper limit for the
validity of Eq.~(\ref{eps}) for the effective permittivity, but it
also sets an upper limit for the frequency band where the
effective permittivity can have negative values in practice.

The existence of a backward wave  can be verified numerically by
plotting the electric field inside a material layer in the
frequency band where the effective medium theory predicts both
material parameters to be negative. Numerical simulations were
made using the finite element method based on Agilent HFSS
electromagnetic modeling software. To test the accuracy of the
solution, a single sphere width radius $r=r_e=3.18$ mm or
$r=r_m=2.28$ mm and $\epsilon_i=44(1-j10^{-4})$ was placed at the
center of an air filled rectangular waveguide. The height and
width of the waveguide were $w=h=15$ mm, and the length $l=15$ mm.
Boundary conditions were ideal electric conductor (PEC) at the top
and bottom of the waveguide and ideal magnetic conductor (PMC) on
the sides of the waveguide. The incident electric field was
vertically polarized. The scattering matrix element $|S_{11}|$ was
solved using about 37\,000 unknowns. The result is shown in
Fig.~\ref{S11}. Similar peaks as in the analytical solution can be
seen, but the numerically calculated resonant frequencies are
slightly higher. When the number of unknowns was increased to
120\,000, then the resonant frequencies were the same as in the
analytical solution. It appears that when using smaller number of
unknowns, the numerical solution overestimates the resonant
frequencies about 1 \%. Both the numerical and analytical studies
show, that spheres can be modeled using frequency dependent
polarizabilities (\ref{alpha}, \ref{beta}) up to the frequency,
where the third resonant mode appears.

\begin{figure}[h!]
\begin{center}
\input{fig7}
\caption{Two and four layers of spheres with similar cross section in the ($x,y$)-plane  were studied. There are four quarter of spheres
in each layer, two with radius $r_1=2.28$ mm and two with radius $r_2=3.18$ mm.}
\label{calc_domain2}
\end{center}
\end{figure}
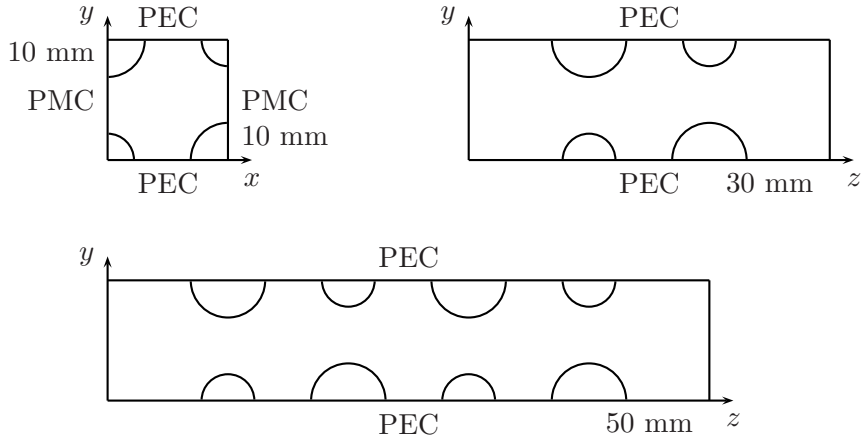

\begin{figure}
\centering
\subfigure[S$_{11}$, $r_1=2.28$ mm and $r_2=3.18$ mm]{\label{slabS11}\includegraphics[width=0.49\linewidth]{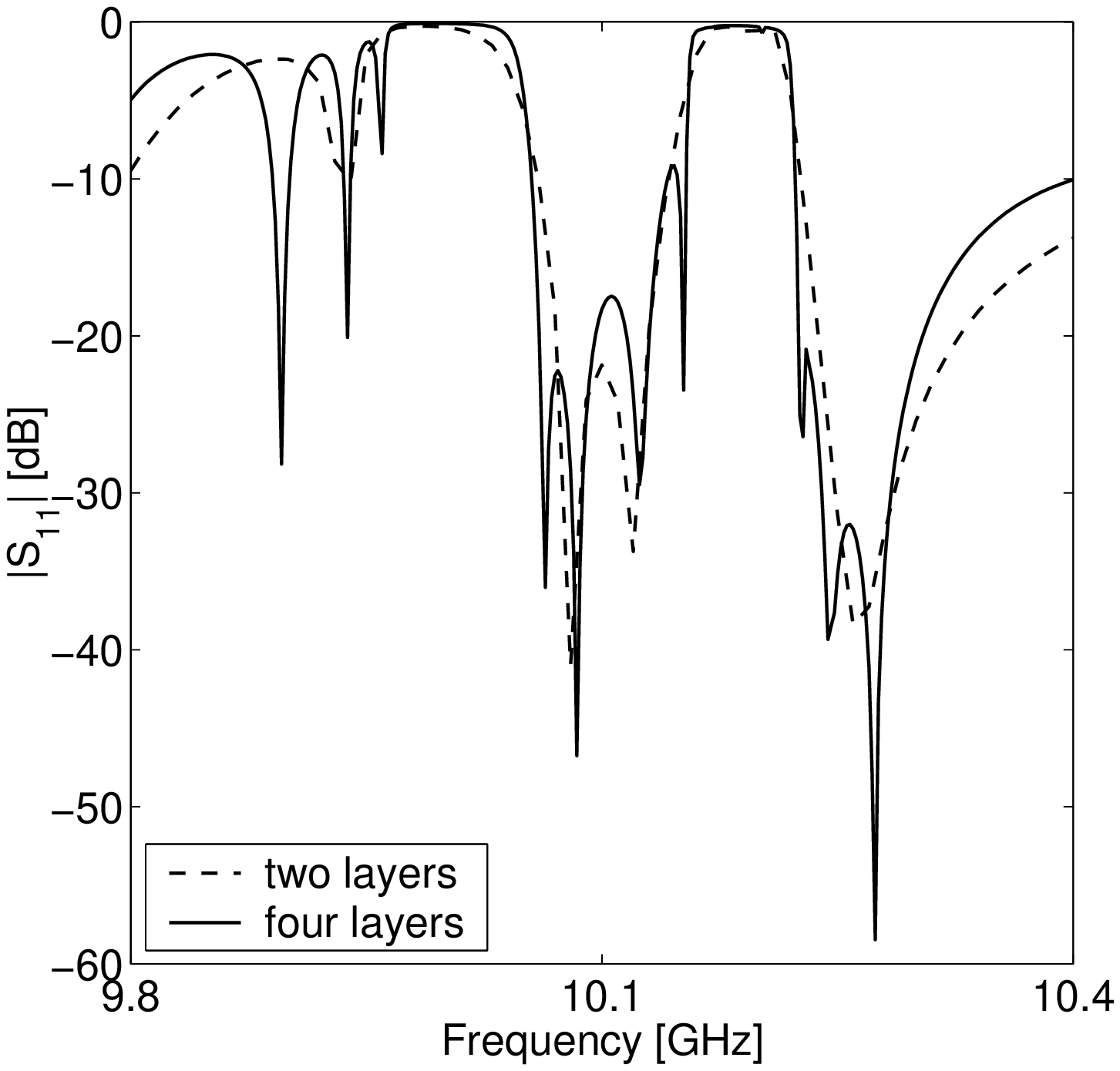}}
%\hspace{2cm}
\subfigure[S$_{21}$, $r_1=2.28$ mm and $r_2=3.18$
mm]{\label{slabS21}\includegraphics[width=0.49\linewidth]{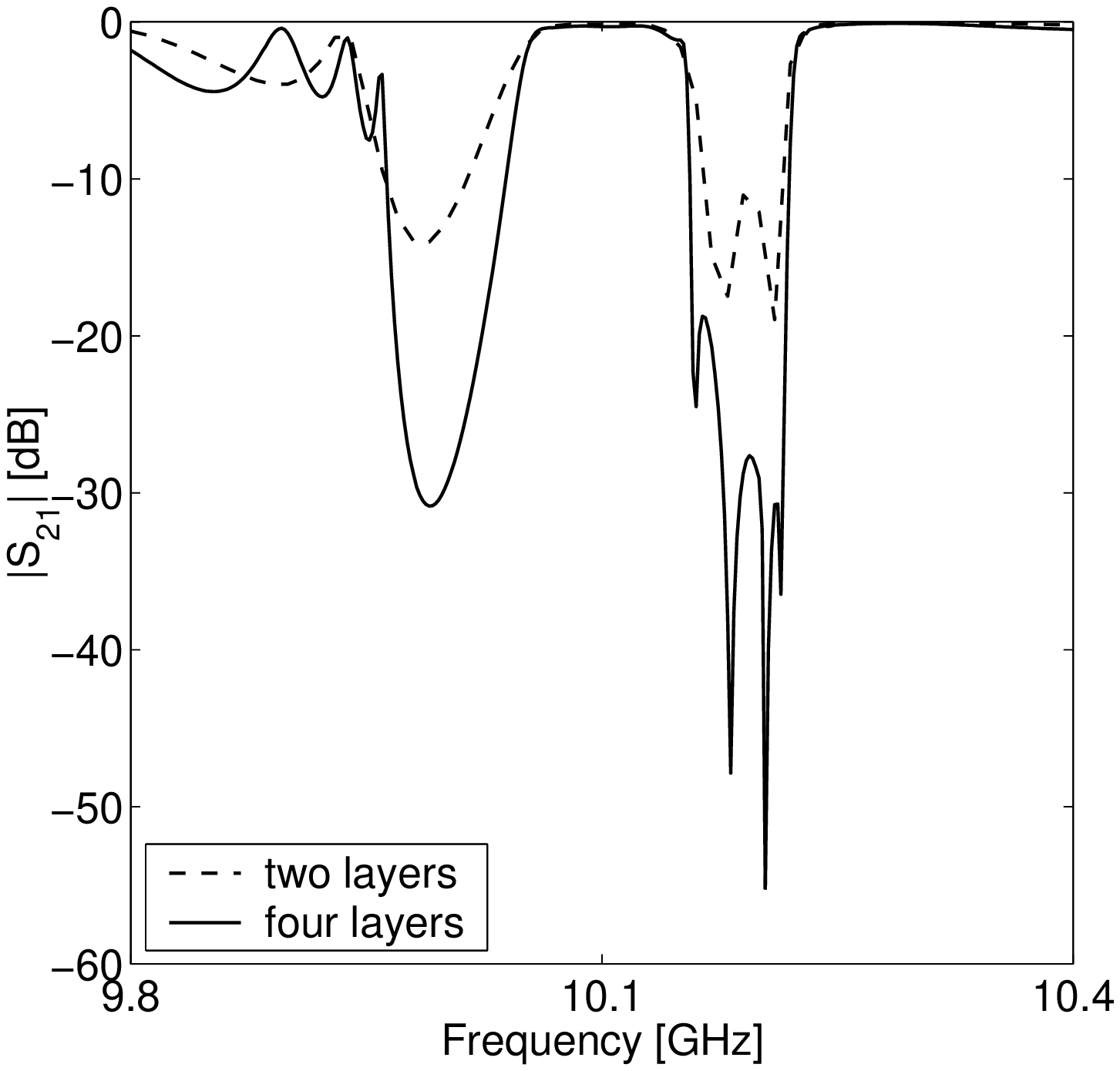}}
\caption{Numerically calculated $S$-parameters for a slab
consisting of two and  four layers of spheres.}
\end{figure}

A material layer consisting of both sets of spheres was modeled
numerically. The accuracy was set to correspond to the simulation
with one sphere with 37\,000 unknowns, accordingly we should
expect numerical simulations to have a frequency shift of about
1\% upwards. An illustration of the calculation domain is shown in
Fig.~\ref{calc_domain2}. Two cases were studied: two and four
layers of spheres.  The studied waveguide corresponds to two or
four infinite layers of spheres, where the nearest neighbors of
spheres with radius $r_e$ are spheres with radius $r_m$ and vice
versa. In the simulations, mirror planes cut spheres into four
parts. These boundary conditions do not disturb the first three
resonant modes of the spheres, which have the same symmetry. The
material parameters and radii were the same as for simulations
with a single sphere in Fig.~\ref{S11}. The results for scattering
matrix elements $|S_{11}|$ and $|S_{21}|$ are shown in
Fig.~\ref{slabS11} and Fig.~\ref{slabS21}. The effective medium
model (\ref{eps}, \ref{mu})  predicts, that this medium should
have negative material parameters from 9.92 GHz to 9.98 GHz. When
the frequency shift caused by numerical error is taken into
account, the negative material parameters should occur around
10.05 GHz. From figures it can be seen, that around this frequency
there is a propagating mode. The stop band after this pass band is
because of the third resonant mode of the larger spheres. If both
material parameters have negative values, the wave inside the
material should be a backward wave. However, it can not be seen
from $S$-parameter studies if the wave is a forward or a backward
wave.

\begin{figure}
\centering
\includegraphics[width=\linewidth]{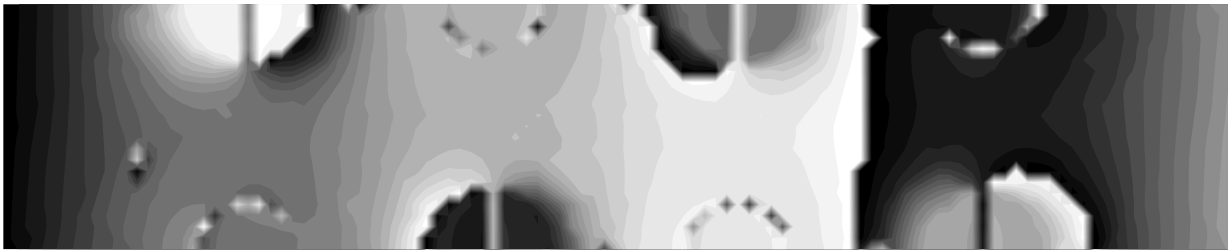}
\caption{The phase of the electric field component $E_y$ at 10.064
GHz for a structure as in Figs.~\ref{slabS11}, \ref{slabS21} with
four layers of spheres. The feeding plane is on the left, and the
energy propagates from left to right. The black color corresponds
to $-\pi$ and the white  to $+\pi$. The phase is increasing as the
distance from the source increases, which indicates that the wave
is a backward wave.} \label{phase}
%\hspace{2cm}
\end{figure}

The existence of a backward wave near 10.05 GHz can be verified
numerically by plotting the electric field distribution inside a
material layer. For the chosen harmonic time dependence the phase
of the electric field component $E_y$ decreases in the positive
direction of the wave vector $k$. Fig.~\ref{phase} displays the
phase of the electric field at 10.064 GHz inside the simulated
waveguide. Black corresponds to $-\pi$ and white to $\pi$. The
feeding point is on the left, so the energy propagates from left
to right, but we see that the phase increases from the left to
right. This clearly shows that this is a backward wave. From
similar field plots it is seen, that the wave is a backward wave
in the entire frequency band around 10.1 GHz. According to
the $S$-parameter studies there is a propagating mode
[Figs.~\ref{slabS11}, \ref{slabS21}]. In the numerical simulations,
the center frequency is about 10.1 GHz and the bandwidth is 1\%.
The center frequency calculated using the effective medium theory
is 9.95 GHz, which corresponds to 10.05 GHz in Fig.~\ref{slabS11}
and \ref{slabS21} because of the numerical overestimation of the
resonant frequencies by about 1\%. The bandwidth in the effective
medium theory is about 0.6\%. We can conclude that the effective
medium model well predicts the frequencies where the backward wave
exists. It appears that the bandwidth in the numerical simulations
is even larger than the prediction of the effective medium model.

%Inside a homogenous material, the phase as a function of the distance from the
%opening of the waveguide would be linear.

\section{Conclusions}

In this study, the possibility to realize an isotropic material
with negative real parts of both $\epsilon$ and $\mu$ using a
mixture consisting of two sets of resonating dielectric spheres
with different radii was considered. A corrected effective medium
model for the effective permittivity was presented, which can be
used  also to calculate how a statistical size distribution
increases the losses in the frequency band where both material
parameters have negative values. The limitations of the analytical
model were discussed and numerically validated.

According to numerical studies, a thin material slab supports a
backward wave in the same frequency band which the effective
medium model predicts. The backward-wave regime corresponds to
both $\epsilon$ and $\mu$ being negative.

In the simulations, quite low contrast between the inclusions and
the environment was used. The introduced effective medium theory for resonant spheres should be
even more applicable when the contrast is larger. For higher frequencies,
the
medium starts to behave as an electromagnetic band gap structure
(EBG), where backward waves can also exist. However, EBG
structures are not isotropic and they can not be modeled using
effective medium theories.

\subsection*{Acknowledgements}

This work has been coordinated and partially funded by the {\it
Metamorphose} Network of Excellence. Financial support  of the
Academy of Finland and TEKES through the Center-of-Excellence
program is acknowledged. Contribution of I. Kolmakov has been
partially funded by an INTAS Young Scientist Fellowship grant and contribution of L. Jylh\"{a} by Finnish Academy of Science and Letters, Vilho, Yrj\"{o} and Kalle V\"{a}is\"{a}la Foundation.

%___________________________________________________________________________________________________________________________________
%***********************************************************************************************************************************
%***********************************************************************************************************************************
%___________________________________________________________________________________________________________________________________
\bibliography{kirjat}

\newpage
%1
%2
%3
%4
%5

%6
%7

\pagestyle{plain}

%_________________________________________________________________

\end{document}

%% file: fig1.tex
\begin{pspicture}(0,0)(10,4)
\newcommand{\alue}[1]{
{\psset{unit=#1}
\pspolygon[linecolor=black,linewidth=0.5mm](0,0)(5,0)(5,5)(0,5)(0,0)
\pspolygon[linecolor=black,linewidth=0.5mm](5.05,0.05)(7.5,2.5)(7.5,7.5)(5,5)(5,0)
\pspolygon[linecolor=black,linewidth=0.5mm](0,5)(2.5,7.5)(7.5,7.5)(5,5)(0,5)
%\psline[linecolor=black,linewidth=0.5mm](0,5)(2.5,7.5)
%\psline[linecolor=black,linewidth=0.5mm](2.5,7.5)(7.5,7.5)
\pscircle[fillstyle=solid,linecolor=black,fillcolor=black](0.5,0.5){0.3}
\pscircle[fillstyle=solid,linecolor=black,fillcolor=black](1.5,0.5){0.2}
\pscircle[fillstyle=solid,linecolor=black,fillcolor=black](2.5,0.5){0.3}
\pscircle[fillstyle=solid,linecolor=black,fillcolor=black](3.5,0.5){0.2}
\pscircle[fillstyle=solid,linecolor=black,fillcolor=black](4.5,0.5){0.3}
\pscircle[fillstyle=solid,linecolor=black,fillcolor=black](0.5,1.5){0.2}
\pscircle[fillstyle=solid,linecolor=black,fillcolor=black](1.5,1.5){0.3}
\pscircle[fillstyle=solid,linecolor=black,fillcolor=black](2.5,1.5){0.2}
\pscircle[fillstyle=solid,linecolor=black,fillcolor=black](3.5,1.5){0.3}
\pscircle[fillstyle=solid,linecolor=black,fillcolor=black](4.5,1.5){0.2}
\pscircle[fillstyle=solid,linecolor=black,fillcolor=black](0.5,2.5){0.3}
\pscircle[fillstyle=solid,linecolor=black,fillcolor=black](1.5,2.5){0.2}
\pscircle[fillstyle=solid,linecolor=black,fillcolor=black](2.5,2.5){0.3}
\pscircle[fillstyle=solid,linecolor=black,fillcolor=black](3.5,2.5){0.2}
\pscircle[fillstyle=solid,linecolor=black,fillcolor=black](4.5,2.5){0.3}
\pscircle[fillstyle=solid,linecolor=black,fillcolor=black](0.5,3.5){0.2}
\pscircle[fillstyle=solid,linecolor=black,fillcolor=black](1.5,3.5){0.3}
\pscircle[fillstyle=solid,linecolor=black,fillcolor=black](2.5,3.5){0.2}
\pscircle[fillstyle=solid,linecolor=black,fillcolor=black](3.5,3.5){0.3}
\pscircle[fillstyle=solid,linecolor=black,fillcolor=black](4.5,3.5){0.2}
\pscircle[fillstyle=solid,linecolor=black,fillcolor=black](0.5,4.5){0.3}
\pscircle[fillstyle=solid,linecolor=black,fillcolor=black](1.5,4.5){0.2}
\pscircle[fillstyle=solid,linecolor=black,fillcolor=black](2.5,4.5){0.3}
\pscircle[fillstyle=solid,linecolor=black,fillcolor=black](3.5,4.5){0.2}
\pscircle[fillstyle=solid,linecolor=black,fillcolor=black](4.5,4.5){0.3}
\psellipse[fillstyle=solid,linecolor=black,fillcolor=black](5.25,0.75)(0.1,0.3)
\psellipse[fillstyle=solid,linecolor=black,fillcolor=black](5.75,1.25)(0.07,0.2)
\psellipse[fillstyle=solid,linecolor=black,fillcolor=black](6.25,1.75)(0.1,0.3)
\psellipse[fillstyle=solid,linecolor=black,fillcolor=black](6.75,2.25)(0.07,0.2)
\psellipse[fillstyle=solid,linecolor=black,fillcolor=black](7.25,2.75)(0.1,0.3)
\psellipse[fillstyle=solid,linecolor=black,fillcolor=black](5.25,1.75)(0.07,0.2)
\psellipse[fillstyle=solid,linecolor=black,fillcolor=black](5.75,2.25)(0.1,0.3)
\psellipse[fillstyle=solid,linecolor=black,fillcolor=black](6.25,2.75)(0.07,0.2)
\psellipse[fillstyle=solid,linecolor=black,fillcolor=black](6.75,3.25)(0.1,0.3)
\psellipse[fillstyle=solid,linecolor=black,fillcolor=black](7.25,3.75)(0.07,0.2)
\psellipse[fillstyle=solid,linecolor=black,fillcolor=black](5.25,2.75)(0.1,0.3)
\psellipse[fillstyle=solid,linecolor=black,fillcolor=black](5.75,3.25)(0.07,0.2)
\psellipse[fillstyle=solid,linecolor=black,fillcolor=black](6.25,3.75)(0.1,0.3)
\psellipse[fillstyle=solid,linecolor=black,fillcolor=black](6.75,4.25)(0.07,0.2)
\psellipse[fillstyle=solid,linecolor=black,fillcolor=black](7.25,4.75)(0.1,0.3)
\psellipse[fillstyle=solid,linecolor=black,fillcolor=black](5.25,3.75)(0.07,0.2)
\psellipse[fillstyle=solid,linecolor=black,fillcolor=black](5.75,4.25)(0.1,0.3)
\psellipse[fillstyle=solid,linecolor=black,fillcolor=black](6.25,4.75)(0.07,0.2)
\psellipse[fillstyle=solid,linecolor=black,fillcolor=black](6.75,5.25)(0.1,0.3)
\psellipse[fillstyle=solid,linecolor=black,fillcolor=black](7.25,5.75)(0.07,0.2)
\psellipse[fillstyle=solid,linecolor=black,fillcolor=black](5.25,4.75)(0.1,0.3)
\psellipse[fillstyle=solid,linecolor=black,fillcolor=black](5.75,5.25)(0.07,0.2)
\psellipse[fillstyle=solid,linecolor=black,fillcolor=black](6.25,5.75)(0.1,0.3)
\psellipse[fillstyle=solid,linecolor=black,fillcolor=black](6.75,6.25)(0.07,0.2)
\psellipse[fillstyle=solid,linecolor=black,fillcolor=black](7.25,6.75)(0.1,0.3)
%
% katon ellipsit:
\psellipse[fillstyle=solid,linecolor=black,fillcolor=black](0.75,5.25)(0.3,0.1)
\psellipse[fillstyle=solid,linecolor=black,fillcolor=black](1.25,5.75)(0.2,0.07)
\psellipse[fillstyle=solid,linecolor=black,fillcolor=black](1.75,6.25)(0.3,0.1)
\psellipse[fillstyle=solid,linecolor=black,fillcolor=black](2.25,6.75)(0.2,0.07)
\psellipse[fillstyle=solid,linecolor=black,fillcolor=black](2.75,7.25)(0.3,0.1)
\psellipse[fillstyle=solid,linecolor=black,fillcolor=black](1.75,5.25)(0.2,0.07)
\psellipse[fillstyle=solid,linecolor=black,fillcolor=black](2.25,5.75)(0.3,0.1)
\psellipse[fillstyle=solid,linecolor=black,fillcolor=black](2.75,6.25)(0.2,0.07)
\psellipse[fillstyle=solid,linecolor=black,fillcolor=black](3.25,6.75)(0.3,0.1)
\psellipse[fillstyle=solid,linecolor=black,fillcolor=black](3.75,7.25)(0.2,0.07)
\psellipse[fillstyle=solid,linecolor=black,fillcolor=black](2.75,5.25)(0.3,0.1)
\psellipse[fillstyle=solid,linecolor=black,fillcolor=black](3.25,5.75)(0.2,0.07)
\psellipse[fillstyle=solid,linecolor=black,fillcolor=black](3.75,6.25)(0.3,0.1)
\psellipse[fillstyle=solid,linecolor=black,fillcolor=black](4.25,6.75)(0.2,0.07)
\psellipse[fillstyle=solid,linecolor=black,fillcolor=black](4.75,7.25)(0.3,0.1)
\psellipse[fillstyle=solid,linecolor=black,fillcolor=black](3.75,5.25)(0.2,0.07)
\psellipse[fillstyle=solid,linecolor=black,fillcolor=black](4.25,5.75)(0.3,0.1)
\psellipse[fillstyle=solid,linecolor=black,fillcolor=black](4.75,6.25)(0.2,0.07)
\psellipse[fillstyle=solid,linecolor=black,fillcolor=black](5.25,6.75)(0.3,0.1)
\psellipse[fillstyle=solid,linecolor=black,fillcolor=black](5.75,7.25)(0.2,0.07)
\psellipse[fillstyle=solid,linecolor=black,fillcolor=black](4.75,5.25)(0.3,0.1)
\psellipse[fillstyle=solid,linecolor=black,fillcolor=black](5.25,5.75)(0.2,0.07)
\psellipse[fillstyle=solid,linecolor=black,fillcolor=black](5.75,6.25)(0.3,0.1)
\psellipse[fillstyle=solid,linecolor=black,fillcolor=black](6.25,6.75)(0.2,0.07)
\psellipse[fillstyle=solid,linecolor=black,fillcolor=black](6.75,7.25)(0.3,0.1)
}}
\newcommand{\ave}[1]{
{\psset{unit=#1}
\pspolygon[linecolor=black,linewidth=0.5mm,fillstyle=solid, fillcolor=lightgray](0,0)(5,0)(5,5)(0,5)(0,0)
\pspolygon[linecolor=black,linewidth=0.5mm,fillstyle=solid, fillcolor=lightgray](5.05,0.05)(7.5,2.5)(7.5,7.5)(5,5)(5,0)
\pspolygon[linecolor=black,linewidth=0.5mm,fillstyle=solid, fillcolor=lightgray](0,5)(2.5,7.5)(7.5,7.5)(5,5)(0,5)
\rput(2.5,3.0){{\Large $\varepsilon_{\rm eff}$}}
\rput(2.5,2.0){{\Large $\mu_{\rm eff}$}}
}}
\newcommand{\kokokuva}[1]{
{\psset{unit=#1}
\rput(0,0){\alue{1}}
\psline[linewidth=1mm]{->}(8,2.5)(11,2.5)
\rput(12,0){\ave{1}}
}}
\rput(0,0){\kokokuva{0.5}}
\end{pspicture}

%% file: fig7.tex
\begin{pspicture}(0,0)(10,7)
\newcommand{\xy}[1]{
{\psset{unit=#1}
% spheres, they are paritially hided so that 1/4 are visible
\pscircle(0,0){0.228}
\pscircle(1,1){0.228}
\pscircle(0,1){0.318}
\pscircle(1,0){0.318}
\pspolygon[fillstyle=solid, fillcolor=white, linecolor=white](-0.4,0)(-0.4,-0.4)(1.4,-0.4)(1.4,0)(-0.4,0)
\pspolygon[fillstyle=solid, fillcolor=white, linecolor=white](-0.4,1)(-0.4,1.4)(1.4,1.4)(1.4,1)(-0.4,1)
\pspolygon[fillstyle=solid, fillcolor=white, linecolor=white](1,-0.4)(1.4,-0.4)(1.4,1.4)(1,1.4)(1,-0.4)
\pspolygon[fillstyle=solid, fillcolor=white, linecolor=white](-0.4,-0.4)(0,-0.4)(0,1.4)(-0.4,1.4)(-0.4,-0.4)
\psline(1,0)(1,1)
\psline(0,1)(1,1)
\psline{->}(0,0)(1.2,0)
\psline{->}(0,0)(0,1.2)
\uput[l](0,1.2){$y$}
\uput[l](0,0.9){\small 10 mm}
\uput[d](1.2,0){$x$}
\uput[r](1,0.2){\small 10 mm}
%\uput[d](0.9,-0.2){\small 10 mm}
\uput[u](0.5,1){\small  PEC}
\uput[d](0.5,0){\small  PEC}
\uput[l](0,0.5){\small  PMC}
\uput[r](1,0.5){\small  PMC}
}}
\newcommand{\yzpieni}[1]{
{\psset{unit=#1}
% spheres, they are paritially hided so that 1/4 are visible
\pscircle(1,0){0.228}
\pscircle(2,1){0.228}
\pscircle(1,1){0.318}
\pscircle(2,0){0.318}
\pspolygon[fillstyle=solid, fillcolor=white, linecolor=white](0.6,0)(0.6,-0.4)(2.4,-0.4)(2.4,0)(0.6,0)
\pspolygon[fillstyle=solid, fillcolor=white, linecolor=white](0.6,1)(0.6,1.4)(2.4,1.4)(2.4,1)(0.6,1)
\psline{->}(0,0)(3.2,0)
\psline{->}(0,0)(0,1.2)
\psline(3,0)(3,1)
\psline(0,1)(3,1)
\uput[l](0,1.2){$y$}
%\uput[l](0,0.9){\small 10 mm}
\uput[d](3.2,0){$z$}
\uput[d](2.5,0){\small 30 mm}
\uput[u](1.5,1){\small  PEC}
\uput[d](1.5,0){\small  PEC}
%\uput[l](0,0.5){\tiny  PMC}
%\uput[r](3,0.5){\tiny  PMC}
}}
\newcommand{\yziso}[1]{
{\psset{unit=#1}
% spheres, they are paritially hided so that 1/4 are visible
\pscircle(1,0){0.228}
\pscircle(2,1){0.228}
\pscircle(1,1){0.318}
\pscircle(2,0){0.318}
\pscircle(3,0){0.228}
\pscircle(4,1){0.228}
\pscircle(3,1){0.318}
\pscircle(4,0){0.318}
\pspolygon[fillstyle=solid, fillcolor=white, linecolor=white](0.6,0)(0.6,-0.4)(4.4,-0.4)(4.4,0)(0.6,0)
\pspolygon[fillstyle=solid, fillcolor=white, linecolor=white](0.6,1)(0.6,1.4)(4.4,1.4)(4.4,1)(0.6,1)
\psline(5,0)(5,1)
\psline(0,1)(5,1)
\psline{->}(0,0)(5.2,0)
\psline{->}(0,0)(0,1.2)
\uput[l](0,1.2){$y$}
%\uput[l](0,0.9){\small 10 mm}
\uput[d](5.2,0){$z$}
\uput[d](4.5,0){\small 50 mm}
\uput[u](2.5,1){\small PEC}
\uput[d](2.5,0){\small PEC}
%\uput[l](0,0.5){\tiny PMC}
%\uput[r](5,0.5){\tiny PMC}
}}
\newcommand{\kokokuva}[1]{
{\psset{unit=#1}
\rput(0,4){\xy{2}}
\rput(6,4){\yzpieni{2}}
\rput(0,0){\yziso{2}}
}}
\rput(0,0.5){\kokokuva{0.8}}
\end{pspicture}